\documentclass[prb,aps,twocolumn,showpacs]{revtex4}
\usepackage{graphicx,latexsym}
\usepackage{dcolumn}
\usepackage{amsmath,amssymb,epsf,bm}
\begin{document}
\title{Edge bands and vertical transport in topological insulator/magnetic insulator  heterostructures}
\author{A.~G. Mal'shukov}
\affiliation{Institute of Spectroscopy, Russian Academy of Sciences,
142190, Troitsk, Moscow oblast, Russia}
\begin{abstract}
The low-energy band-structure of electrons propagating on a lateral surface of a heterostructure consisting of three dimensional topological insulator (TI) and magnetic insulator layers has been calculated. The energy spectrum is highly tunable depending on the relation between the interlayer tunneling amplitude, Zeeman energy and surface potential. A ratio between the first two parameters controls a topological  transition between chiral and helical surface state band-structures. In the former case localized states, whose wave-functions are confined near single TI layers, emerge in a narrow parameter range where they coexist with itinerant states. In regular superlattices these localized states form a flat band in the vertical direction. Such a localization has a topological origin. It is associated with a specific spatial spin texture of these states. The localized states vanish upon the topological phase transition from the anomalous quantum Hall regime to a trivial phase characterized by an anisotropic Dirac cone on the lateral surface of the heterostructure.
\end{abstract}
\maketitle

\section{Introduction}
Three dimensional topological insulators (TI) are band insulators which carry metallic electronic  states on their surfaces. These states are described by the two-dimensional massless Dirac Hamiltonian with an odd number of Dirac points (for a review see Ref. \onlinecite{Qi RMP,Hasan}). The corresponding surface eigenstates are characterized by a helical structure where the spin and momentum are rigidly locked by a spin-orbit coupling. The surface states are robust with respect to perturbations that obey time-reversal symmetry.\cite{Fu} This symmetry can be broken by the exchange field which may be induced by ferromagnetically  ordered  magnetic impurities in TI, \cite{magnetic doping} or by TI contact with a magnetic material. \cite{Wei}  As a result, the effective Zeeman term in the Dirac Hamiltonian produces a gap in the energy spectrum of the surface states, transforming the two-dimensional metal into the Chern insulator. \cite{Qi RMP,Hasan} This phase is characterized by a half-quantized anomalous Hall effect (AHE) generically related to the topological magnetoelectric effect in TI. \cite{Qi Basics TI,Essin}. The latter effect gives rise to a number of interesting magnetoelectric phenomena with many potential applications. \cite{TME}

The anomalous Hall current is carried by  one-dimensional (1D) chiral conducting channels  residing on domain walls between exchange fields of opposite signs \cite{Qi RMP,Hasan,Fu,Qi Basics TI}. These channels can be also localized near a boundary between different crystal facets \cite{Zhang facets}, as well as near 1D edges of an interface between two TI's separated by a thin tunneling barrier, \cite{Meng folding} providing that their surfaces are in the insulating phase due to the Zeeman interaction. The latter two proposals are interesting from the practical point of view, since they suggest a new way to observe the quantized anomalous Hall effect. It is interesting to extend these ideas to heterostructures and superlattices containing TI layers and ordinary-insulator, or magnetic-insulator layers. At the same time, recent studies show that superlattices with desirable properties can be fabricated. Ab initio calculations of electronic states in a Bi$_2$Se$_3$/MnSe superlattice  \cite{calculations for superlattices} show a high tunability of these systems in the important parameter range, where the AHE can be realized. Recently, there were reports on naturally grown TI heterostructures \cite{suplattice Bi2Bi2Se3,natural heterostruc Ando} and superlattices based on intergrown magnetic and TI layers. \cite{Ji intergrowth} To exhibit quantum AHE the former must include magnetic components, that may be achieved, for example, by an appropriate magnetic doping.

The main goal of the present work is to study conducting electron states that carry the anomalous Hall current on the lateral surface of a vertically grown heterostructure  (see Fig. 1). To our knowledge, this problem has not been addressed yet, although in some recent works edge states have been theoretically studied for single TI layers and slabs \cite{Brey,Deb,Zhang facets}, but not for a multilayer system. At the same time,  propagation of electrons in a multilayer system can be quite unusual due to a specific spacial spin texture associated with the chiral character of electron states. As it will be shown in this work, this texture leads to localization of electron's wave-functions at single TI layers, even in the case of a relatively large tunnel coupling between TI layers. In regular superlattices it shows up in formation of flat bands in the vertical direction. Such a localization becomes possible due to a nontrivial topology of electron states on TI surfaces and it has no analogies in other systems, although one may see some resemblance to flat bands in Su-Schrieffer-Heeger one-dimensional model. \cite{Su} On the side surface of a superlattice the localized states coexist with itinerant states. All these states provide conducting channels carrying the anomalous Hall current. In previous studies of edge states in TI films and slabs \cite{Brey,Deb,Zhang facets} in the quantum Hall regime the considered system compositions and  sizes allowed the existence of only a single conducting channel. On the other hand, even a single TI layer can carry several such channels, depending on its thickness. In ideal situation, without scattering it is expected that in the quantum Hall regime the total conductance of these channels is equal to one conductance quantum, while it turns to zero in the trivial phase. We considered such a multichannel case for a single TI layer and for a superlattice and demonstrated the expected reconstruction of edge bands upon a topological transition between the quantum Hall and trivial phases. In turn, the topological phase transition can be controlled by a ratio between the Zeeman interaction and the interlayer tunneling parameter.

The system to be studied  consists of anisotropic TI films separated by tunneling barriers. The electronic surface states at interfaces and lateral faces of these films are described by corresponding Dirac Hamiltonians, with the mass term  generated by the Zeeman field $m$ which is parallel to $z$. In the case when this field is produced by the proximity effect with a magnetic-insulator layer, as in Ref. \onlinecite{calculations for superlattices}, the mass term appears only at interface. In this case, in the AHE regime the entire conducting side surface of each TI film carries a quantum of the Hall current. \cite{Zhang facets} The wave function of an edge state which resides on this surface penetrates into the interface region where  the interlayer tunneling parameter $\Delta$ connects the wave functions of neighboring layers. A final equation for determining the energies and wave-functions of the edge states can be obtained by using the matching conditions at boundaries between horizontal and lateral crystal faces. This equation has been analyzed numerically, as well as analytically in some  limiting cases.

The article is organized in the following way.  In Sec. II the Hamiltonian of the entire heterostructure including horizontal and side faces of anisotropic TI is formulated and the matching conditions are found that  allow to meet wave functions on horizontal and side faces of TI layers. Also the  equations are derived for calculation of the electronic band energies. In Sec.III  we demonstrate for a general heterostructure that  the states localized near single TI layers can exist on its lateral surface at $|m|>\Delta$. The results for a regular superlattice are presented in Sec. IV. The analytical results in some limiting cases are obtained for the topologically distinct regimes $|m|>\Delta$ and $|m|<\Delta$ . We considered the effect of a spin-dependent tunneling on the flat-band wave-functions. Also, in this section numerical results are presented in a broader range of parameters. These results are discussed in Sec. V.

\section{Hamiltonian and wave functions}

\begin{figure}[t]
\includegraphics[width=6cm]{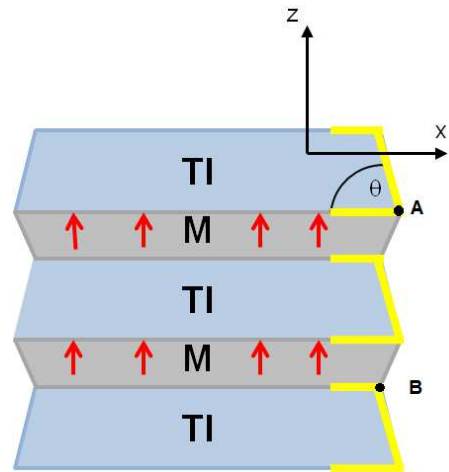}
\caption{(Color online) A heterostructure consists of topological insulator (TI) and magnetic insulator (M) layers. The orientation of the side facets of TI layers is denoted by the angle $\theta$. Two adjacent horizontal TI faces are coupled to each other via electron tunneling through a thin magnetic layer between them. An edge state occupies a region schematically shown in yellow. The wave function of a localized edge state is finite between the points A and B, if it is localized near a central TI layer.  }\label{fig1}.
\end{figure}
Let us consider the n-th element of the superlattice. It consists of a tunneling barrier and two adjacent TI surfaces. We assume that the TI films are thick enough, so that the tunneling between two opposite surfaces of each TI film can be neglected. For layered TI, such as Bi$_2$Se$_3$, this assumption is well justified even for thin films composed of 4 quintuple layers, as it follows from \textit{ab initio} calculation in Ref. \onlinecite{calculations for superlattices}. Each n-th bilayer system contacts at its edges to side surfaces of TI films. At these edges a wave-function of the bilayer must meet with wave functions of adjacent side surfaces. In realistic structures these surfaces are not vertical and their angular orientation depends on film-growth conditions.  In Ref. [\onlinecite{Zhang facets PRB}] a method has been suggested which allows to calculate surface-state wave-functions for an arbitrary crystal face of a three dimensional topological insulator whose crystal lattice belongs to $R3m$ space group. In this method each face is described by the polar $\theta$ and azimuthal $\phi$ angles. Due to azimuthal rotational symmetry (in the linear in $\mathbf{k}$ regime) the $x$-axis can be chosen arbitrary. Therefore, we set $\phi=0$. The polar angles of the upper and bottom surfaces of a TI film  are defined as $\theta=0$ and $\theta=\pi$, respectively. Further, for a facet characterized by the angle $\theta$ the effective low-energy surface Hamiltonian can be written in the form \cite{Zhang facets,Zhang facets PRB}
\begin{equation}\label{Htheta}
H_{\theta}=(v_{\theta}k_{\theta}\sigma_{\theta y} -v_{\parallel}k_{y}\sigma_{\theta x})\tau_{\theta 3}+U_{\theta}+M_{\theta}+v_0k_{\theta}\tau_{\theta 2}
\end{equation}
where $v_{\theta}=v_{\parallel}v_{z}/\sqrt{(v_{z}\cos\theta)^2+(v_{\parallel}\sin\theta)^2}$ and $v_0=(v_{\parallel}^2-v_{z}^2)\sin\theta\cos\theta/v_{\theta}$ are the angle-dependent velocities expressed in terms of the bulk TI velocities $v_{\parallel}$ and $v_{z}$ \cite{Zhang facets PRB}; $k_{\theta}$ is the wave-vector tangential to the face in a rotated $xz$-frame. The spin operators $\sigma_{\theta i}$ ($i=x,y,z$) can be represented as $\sigma_{\theta i}=\mathrm{R}(\theta)\sigma_{i}\mathrm{R}(\theta)^{-1}$, where $\sigma_{i}$ are Pauli matrices and the unitary transformation $R$ is given by
\begin{equation}\label{R}
\mathrm{R}(\theta)=\exp(-i\frac{\nu}{2}\sigma_y\tau_1),
\end{equation}
where $\tau_1,\tau_2$ and $\tau_3$ are Pauli matrices operating in the orbital space. The angle $\nu$ is determined by $\cos\nu=v_z \cos\theta/\sqrt{(v_{z}\cos\theta)^2+(v_{\parallel}\sin\theta)^2}$. According to Ref. [\onlinecite{Zhang facets PRB}], the surface state at a facet is characterized by a positive eigenvalue of $\tau_{\theta 3}=\mathrm{R}(\theta)\tau_{3}\mathrm{R}(\theta)^{-1}=\cos\nu\tau_3-\sin\nu\sigma_y\tau_2$. \cite{redefine} This condition is necessary to get a proper (evanescent in the bulk of TI) state. For example, the top ($\theta$=0) and bottom ($\theta=\pi$) surfaces of each TI layer are represented by $\tau_3=1$ and $\tau_3=-1$, respectively.

$U_{\theta}$ is the face-dependent surface potential (an energy shift of the Dirac point\cite{Zhang facets PRB,Silvestrov}) which is measured relative $U_{0}$ on the upper face of each TI film.  Besides the inbuilt potential, $U_{\theta}$ can also include extrinsic terms of various physical origin. \cite{Zhang facets PRB} Since the side surface is a 2D metal, the screening effects might screen-out the potential difference between horizontal and lateral faces. However, for thin TI layers ($<$10 nm) such a screening is not efficient, at least in the case of Bi$_2$Se$_3$ based materials, where the corresponding screening length is $\sim$ 100nm. \cite{Silvestrov} For simplicity, it will be assumed that $U_{0}=U_{\pi}$, although in reality for topmost and bottom faces of a heterostructure these potentials can differ because of different interface materials. In fact, they may  differ even for internal layers of a superlattice due to asymmetry of the epitaxy-film growing process. $M_{\theta}$ is the Zeeman interaction, which is assumed to be presented only on horizontal faces, where $\theta=0$, or $\theta=\pi$.  For these faces the exchange fields have the same directions and magnitudes, so that in the rotated coordinate frame $M_0=-M_{\pi}=m\sigma_z$.

It is important to note that Hamiltonian (\ref{Htheta}) is valid only in the range of electron energies close to a Dirac point. It contains phenomenological parameters, such as scalar and magnetic potentials $U_{\theta}$ and $M_{\theta}$, which can be calculated by solving a three-dimensional problem of TI interface with magnetic, or other materials. Besides $U$ and $M$, potentials of various symmetries can appear in (\ref{Htheta}), depending on a specific interface. They belong to $SU(2)\otimes SU(2)$ manifold associated with spin and orbital quantum numbers. These potentials are classified in  Refs. \onlinecite{Zhang facets PRB,Zhang facets}. Within the low-energy expansion they are assumed to be much less than the bulk gap, so that they can be treated as perturbations to the TI-vaccum interface. \cite{Zhang facets PRB,Zhang facets} In practice, however, the effect of interface on TI can be very strong. For example, a contact of Bi$_2$Se$_3$  with the magnetic insulator  MnSe leads to a large band bending within the first quintuple layer of TI. \cite{calculations for superlattices, Eremeev} Such a shift is associated with a strong ionic polarization of the MnSe crystal lattice. As it follows from DFT calculations, \cite{Eremeev} due to the band bending the surface state which is localized in the near vicinity of the interface is shifted downward, below the top of  TI valence band. At the same time, one more surface state emerges within the bulk gap. This state is mostly localized in the second quintuple layer. Its spectrum looks as a Dirac cone with a mass term $\sim$ 4 meV.  This mass term  is related to magnetism of MnSe in a rather complicated manner. However, symmetry arguments  do not suggest much choice \cite{Zhang facets} for the cleavage surface, but the mass term in the form $m\sigma_z$ in Eq. (\ref{Htheta}).  Besides this term, six types of  potentials that preserve the time reversion symmetry have been analyzed in Ref. \onlinecite{Zhang facets PRB}. It is argued that in the phenomenological low-energy Hamiltonian their effect is reduced to a shift $U_{\theta}$ of the Dirac point. For  Bi$_2$Se$_3$/MnSe system one can not expect that face-to-face variations of $U_{\theta}$ are much smaller than the bulk gap. Hence,  the low-energy phenomenology can not be applied on the whole surface of TI.  On the other hand, the shift of the Dirac point can be tuned by various adsorbed species.\cite{Zhang facets PRB} Also, the strong electrostatic effect from MnSe and similar magnetic insulators can be diminished by a thin spacer layer. \cite{calculations for superlattices} Probably, other magnetic materials can be found that do not lead to the so strong band-bending effect. Anyway, it will be assumed below that the magnetic gap and the Dirac point variations are small enough, so that Hamiltonian (\ref{Htheta}) can be employed for calculation of edge states.

In general, the wave-function $\psi$ of the surface state is represented by a four-dimensional vector in the space of coupled spin and orbital variables. The unitary transformation
\begin{equation}\label{RPsi}
\psi=\mathrm{R}(\theta)\Psi
\end{equation}
effectively decouples these variables. Indeed, in the new basis Hamiltonian (\ref{Htheta}) keeps its original form with $\sigma_{\theta y}\rightarrow\sigma_y$ and $\sigma_{\theta x}\rightarrow\sigma_x$, and $M_0=-M_{\pi}=m\sigma_z$. So, these operators do not contain orbital variables. At the same time, the equation $\tau_{\theta 3}\psi=\psi$, which filters the proper surface state (evanescent in the bulk of TI), transforms to the angular-independent equation $\tau_{3}\Psi=\Psi$. Hence, for all faces the $\Psi$-vector is fixed in the pseudospin space, while it is an arbitrary spinor  in the spin space. For each facet this spinor can be found from transformed  Hamiltonian (\ref{Htheta}), which becomes an effectively 2$\times$2 matrix when the last term in Eq.  (\ref{Htheta}) is ignored. This term can be ignored on flat facets, as it was discussed in Ref. \onlinecite{Zhang facets PRB}. Nevertheless, we will see that it becomes important in the range of a boundary between facets with different $\theta$.

\emph{\textbf{Matching conditions}}. In order to find  wave functions and energies of modes that propagate on side facets,  one needs matching conditions on the lines connecting neighboring faces.  The matching condition can be easy established for a smooth boundary between two faces characterized by $\theta_1$ and $\theta_2$. Let us assume that the curvature radius $r$ of a surface connecting these faces is much larger than the penetration length of the surface state into TI bulk. At the same time, $r$ is much less than the characteristic geometric sizes of a TI layer. In this case, on the curved surface Eq. (\ref{Htheta}) is valid, where $\theta$ is a function of the local coordinate $l$ along the line connecting two facets and $k_{\theta}=-i\partial/\partial l$. Applying unitary transformation (\ref{R}) and taking into account only rapidly varying terms in Hamiltonian (\ref{Htheta}), the Schr\"{o}dinger equation projected onto subspace $\tau_3=1$ takes the form
\begin{equation}\label{bc1}
-iv_{\parallel}v_z \sigma_y\left(\frac{1}{v_{\theta}}\frac{\partial \Psi}{\partial l}+\frac{1}{2}\frac{\partial v_{\theta}^{-1}}{\partial l}\Psi\right)=0.
\end{equation}
This equation means that in the boundary region $\Psi$ scales as $\sqrt{v_{\theta}}$. Hence, we obtain the matching condition
\begin{equation}\label{bc}
\sqrt{v_{\theta_1}}\Psi(\theta_1)=\sqrt{v_{\theta_2}}\Psi(\theta_2)
\end{equation}
which guarantees the current conservation across the boundary. In the case of a sharp interface between $\theta=0$ and $\theta=\pi/2$ facets this matching condition coincides (after unitary transformation (\ref{RPsi})) with that obtained in Ref.\onlinecite{Brey,Deb}.

\emph{\textbf{Wave functions of edge states on horizontal faces}}. For n-th bilayer system consisting of top and bottom interfaces of TI layers with a magnetic tunneling barrier in between,  we can consider  the wave-function $\psi_n(x,y)$ as a two-component  vector $(\psi_{n1},\psi_{n2})$, where $\psi_{n1}$ and $\psi_{n2}$ are spinors defined on the bottom and top interfaces, respectively. On the other hand, $\psi_{n1}$ corresponds to $\theta=0$, and its orbital pseudospin $\tau_3=1$, while $\psi_{n2}$ is associated with $\tau_3=-1$. Hence, at a given energy $E$ the wave function of n-th bilayer satisfies the equation $H\psi_n=E\psi_n$, where, according to Eq.(\ref{Htheta})
\begin{equation}\label{H}
H=v_{\parallel}(\hat{k}_{x}\sigma_{y} -\hat{k}_{y}\sigma_{x})\tau_3 + \Delta\tau_1 +m\sigma_z\,.
\end{equation}
The Pauli matrices $\tau_1,\tau_2,\tau_3$ operate in the two-component space of $\psi_n$. The real term $\Delta \tau_1$ is added to (\ref{H}) in order to take into account a tunnel coupling of two TI layers.  Without loss of generality $\Delta$ can be chosen positive. This tunneling term preserves the time reversion symmetry $\mathcal{T}$. Within the phenomenological approach three other tunneling terms are possible: $\tau_2$ and $\tau_1\sigma_z$ that break $\mathcal{T}$, and $\mathcal{T}$-invariant $\tau_2\sigma_z$. The former term has no reasonable physical origin in the considered system. Therefore, it will be ignored. Two other terms describe spin-dependent tunneling through a magnetic layer whose magnetic moment is parallel to $z$. We will consider $\gamma\tau_1\sigma_z$ as a perturbation to $\Delta \tau_1$.

Unlike Eq.(\ref{Htheta}), Eq.(\ref{H}) is written in the fixed $x,y,z$ frame. Therefore, the local frame of the top surface in Eq. (\ref{H}) (which is the bottom face of a TI layer in Eq.(\ref{Htheta})) has been rotated by $\pi$ around the $y$-axis. It is convenient to unitary transform Eq.(\ref{H}) according to $H\rightarrow U^{-1}HU$ and $\psi \rightarrow U^{-1}\psi$, where $U=(1/\sqrt{2})(1-i\tau_2\sigma_z)$. After this transformation Eq.(\ref{H}) becomes diagonal in the $\tau$-space:
\begin{equation}\label{H2}
U^{-1}HU=v_{\parallel}(\hat{k}_{x}\sigma_{y} -\hat{k}_{y}\sigma_{x})\tau_3 + \Delta\tau_3\sigma_z +m\sigma_z\,.
\end{equation}
This Hamiltonian describes two massive Dirac particles. In the following it will be assumed that the Fermi-level is placed within their mass gaps $|m\pm\Delta|$. In this case at $m>\Delta$ the system is a Chern insulator with the quantized anomalous Hall conductivity, while at $m<\Delta$ it is a spin-Hall insulator, \cite{Meng folding,film Hamiltonian} where the role of "spin" is played by $\tau_3$. In both cases the Hall and spin-Hall currents are provided by the conducting edge of the system, which in our case is an entire side surface of a  heterostructure built of TI and magnetic layers.  The corresponding surface states are expected to decay in the bulk of a two-dimensional system described by Hamiltonian (\ref{H2}). We assume that the bilayer system occupies the region $x<0$ (see Fig. 1). Its edge is placed at $x=0$ and for thin magnetic-insulator layers one can neglect a relative shift of the edge positions on the top and bottom interfaces. Hence, at the fixed energy $E$ we look for $\psi_n(x,y)$ that exponentially decreases  at $x<0$  and is the plane-wave in the $y$-direction. Such a solution has the form
\begin{equation}\label{psi}
\psi_n(x,y)=\exp(\kappa x)\exp(ik_yy)\psi_n(k_y).
\end{equation}
With this ansatz  the two normalized eigenvectors of Hamiltonian (\ref{H2}), with positive $\kappa$, can be represented as
\begin{align}\label{psi1}
\psi^{(1)}(k_y)=
\begin{pmatrix}
\cos\phi_1\\
\sin\phi_1
\end{pmatrix}
\,\,, \,\tau_3=1
\end{align}
and
\begin{align}\label{psi2}
\psi^{(2)}(k_y)=
\begin{pmatrix}
\cos\phi_2\\
-\sin\phi_2
\end{pmatrix}
\,\,, \,\tau_3=-1.
\end{align}
The parameters $\kappa$ in Eq.(\ref{psi}) are  given, respectively, by
\begin{equation}\label{kappa}
\kappa_{1(2)}=\frac{1}{v_{\parallel}}\sqrt{(m\pm\Delta)^2+v_{\parallel}^2k_y^2-E^2},
\end{equation}
where the "+" and "-" signs in $\pm$ relate to $\kappa_{1}$ and $\kappa_{2}$, respectively. The angles $\phi_1$ and $\phi_2$ in Eq.(\ref{psi1}-\ref{psi2}) are given by
\begin{eqnarray}\label{phi12}
\cos\phi_{1(2)}&=& v_{\parallel}(\kappa_{1(2)}+k_y)/N_{1(2)} \nonumber \\
\sin\phi_{1(2)}&=&(m\pm\Delta-E)/N_{1(2)}\,,
\end{eqnarray}
where
\begin{equation}\label{N}
N_{1(2)}=\sqrt{(m\pm\Delta-E)^2+v_{\parallel}^2(\kappa_{1(2)}+k_y)^2}
\end{equation}

A general solution is given by a linear combination of vectors (\ref{psi}) corresponding to the two eigenvectors (\ref{psi1}) and (\ref{psi2}). This wave-function must match with the functions on the side surface at  $x=0$. At this position the linear combination is
\begin{equation}\label{AB}
\psi_n(k_y)=A_n\psi^{(1)}(k_y)+B_n\psi^{(2)}(k_y)\,.
\end{equation}
By the unitary transformation $U\psi_n(k_y)$ the wave-function (\ref{AB}) is returned to its original representation where it has the form
\begin{eqnarray}\label{psifin}
\psi_n(k_y)&=&\left(A_n\psi^{(1)}(k_y)-B_n\tilde{\psi}^{(2)}(k_y)\right)\delta_{\tau_3,1}+  \nonumber \\
&&\left(A_n\tilde{\psi}^{(1)}(k_y)+B_n\psi^{(2)}(k_y)\right)
\delta_{\tau_3,-1}\,,
\end{eqnarray}
where $\tilde{\psi}^{(1)}_{1}=\sigma_z\psi^{(1)}_{1}$ and $\tilde{\psi}^{(2)}_{2}=\sigma_z\psi^{(2)}_{2}$.
The Kronecker symbols in the right-hand side of Eq.(\ref{psifin}) mean that the first and the second terms are related to the bottom and top surfaces, respectively. Eq. (\ref{psifin}) can be written in a more compact form by introducing a vector
\begin{align}\label{chin}
\chi_{n}=
\begin{pmatrix}
A_n \\
B_n
\end{pmatrix}
\end{align}
and two matrices
\begin{align}\label{MtMb}
\mathrm{M}_t=
\begin{pmatrix}
\cos\phi_{1}& -\cos\phi_{2}\\
\sin\phi_1& -\sin\phi_2
\end{pmatrix}
\,;\,\mathrm{M}_b=
\begin{pmatrix}
\cos\phi_{1}& \cos\phi_{2}\\
-\sin\phi_1&-\sin\phi_2
\end{pmatrix}
.
\end{align}
By using Eqs. (\ref{chin}) and  (\ref{MtMb}) $\psi_n(k_y)$ is reduced to
\begin{equation}\label{psifin2}
 \psi_n(k_y)=(\mathrm{M}_t  \delta_{\tau_3,1}+ \mathrm{M}_b  \delta_{\tau_3,-1})\chi_n
\end{equation}

\emph{\textbf{Wave functions of edge states on a lateral facet}}. Since the first and the second terms in Eq. (\ref{psifin2}) are associated with the top and bottom planes of n-th bilayer, these terms enter into the matching conditions with wave-functions of adjacent side faces of the TI films in a heterostructure.  Let us now consider the wave-function on a side facet which is characterized by an arbitrary angle  $\theta$. In the adopted model the Dirac Hamiltonian of surface electrons on this face does not contain the mass term. Hence, in Eq.(\ref{Htheta}) one can set $M_{\theta}=0$. After applying unitary transformation (\ref{R}) to Eq. (\ref{Htheta}) and projecting onto the Hilbert subspace $\tau_3=1$, the Hamiltonian of the side face becomes
\begin{equation}\label{Hs}
\mathrm{R}^{-1}(\theta)H_{\theta}\mathrm{R}(\theta)=v_{\theta}\hat{k}_{\theta}\sigma_{y} -v_{\parallel}k_{y}\sigma_{x}+U_{\theta}\,.
\end{equation}
At a given energy $E$ its eigenstates are plane waves
\begin{align}\label{psis}
\psi_{\pm}(x_{\theta})=\frac{1}{\sqrt{2}}e^{\pm ikx_{\theta}}
\begin{pmatrix}
e^{\pm i\phi}\\
e^{\mp i\phi}
\end{pmatrix}
\,,
\end{align}
where
\begin{equation}\label{kphi}
kv_{\theta}=\sqrt{(E-U_{\theta})^2-k^2_yv_{\parallel}^2}\,\, ;\,\,e^{2i\phi}=\frac{U_{\theta}-E}{v_{\parallel}k_y+iv_{\theta}k}\,.
\end{equation}
The local coordinate $x_{\theta}$ varies from 0 to $L$, where the side face meets, respectively, with the top and bottom horizontal planes of the TI layer. The general solution $\psi_{\theta}(x_{\theta})$ is given by a linear combination of two functions Eq. (\ref{psis}). By introducing the matrices
\begin{align}\label{MT}
\mathrm{M}_s=\frac{1}{\sqrt{2}}
\begin{pmatrix}
e^{i\phi}&e^{-i\phi} \\
e^{-i\phi}&e^{i\phi}
\end{pmatrix}
\,\,\,\, \text{and}\,\,\,\,\mathrm{T}=
\begin{pmatrix}
e^{ikL}&0\\
0&e^{-ikL}
\end{pmatrix}
\end{align}
the wave-function at $x_{\theta}=L$ can be expressed through $\psi_{\theta}(0)$ according to
\begin{equation}\label{psiL}
\psi_{\theta}(L)=\mathrm{M}_s\mathrm{T}\mathrm{M}_s^{-1}\psi_{\theta}(0)\,.
\end{equation}
In general, the parameters entering into this equation, such as $L$, $\theta$, $U_{\theta}$, can vary from layer to layer.

Now one can use Eq.(\ref{bc}) to match the wave-functions $\psi_{\theta}(L)$ and $\psi_{\theta}(0)$ with $\psi_n$ given by Eq. (\ref{psifin2}). Specifically, the first term in $\psi_{n+1}$ meets with $\psi_{\theta}(0)$, while $\psi_{\theta}(L)$ meets with the second term in $\psi_{n}$. The latter term originating from the bottom surface of a TI film must be preliminary rotated by $\mathrm{R}(\pi)=-i\sigma_y$. As a result, one arrives to the equation
\begin{equation}\label{maineq}
\mathrm{M}_s\mathrm{T}\mathrm{M}_s^{-1}\mathrm{M}_t\chi_{n+1}-
\mathrm{R}(\pi)\mathrm{M}_b\chi_{n}=0\,.
\end{equation}
This equation determines the energy spectrum and eigenvectors  $\chi_n$ of edge states in a heterostructure.
\begin{figure}[t]
\includegraphics[width=6cm]{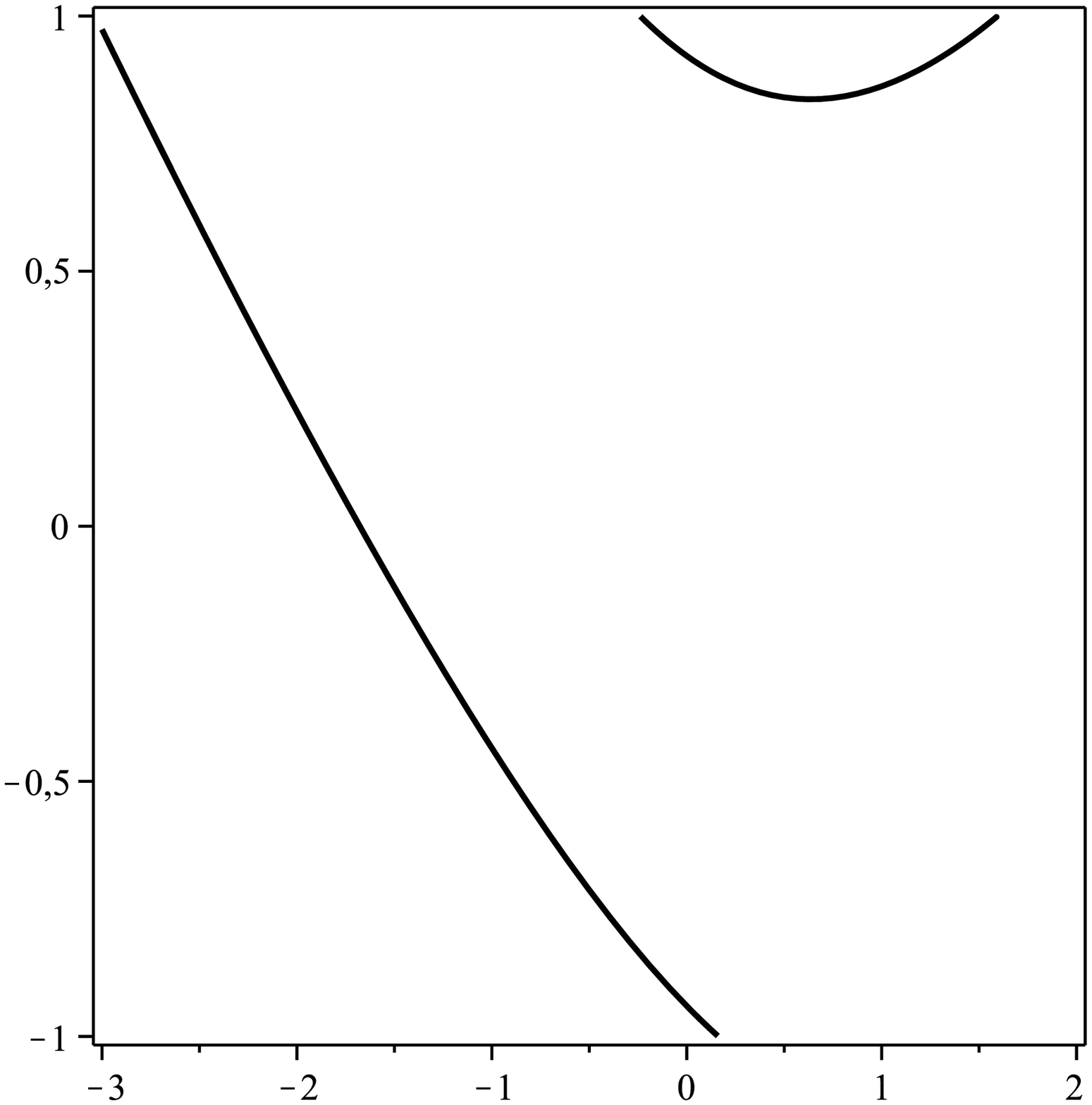}
\includegraphics[width=6cm]{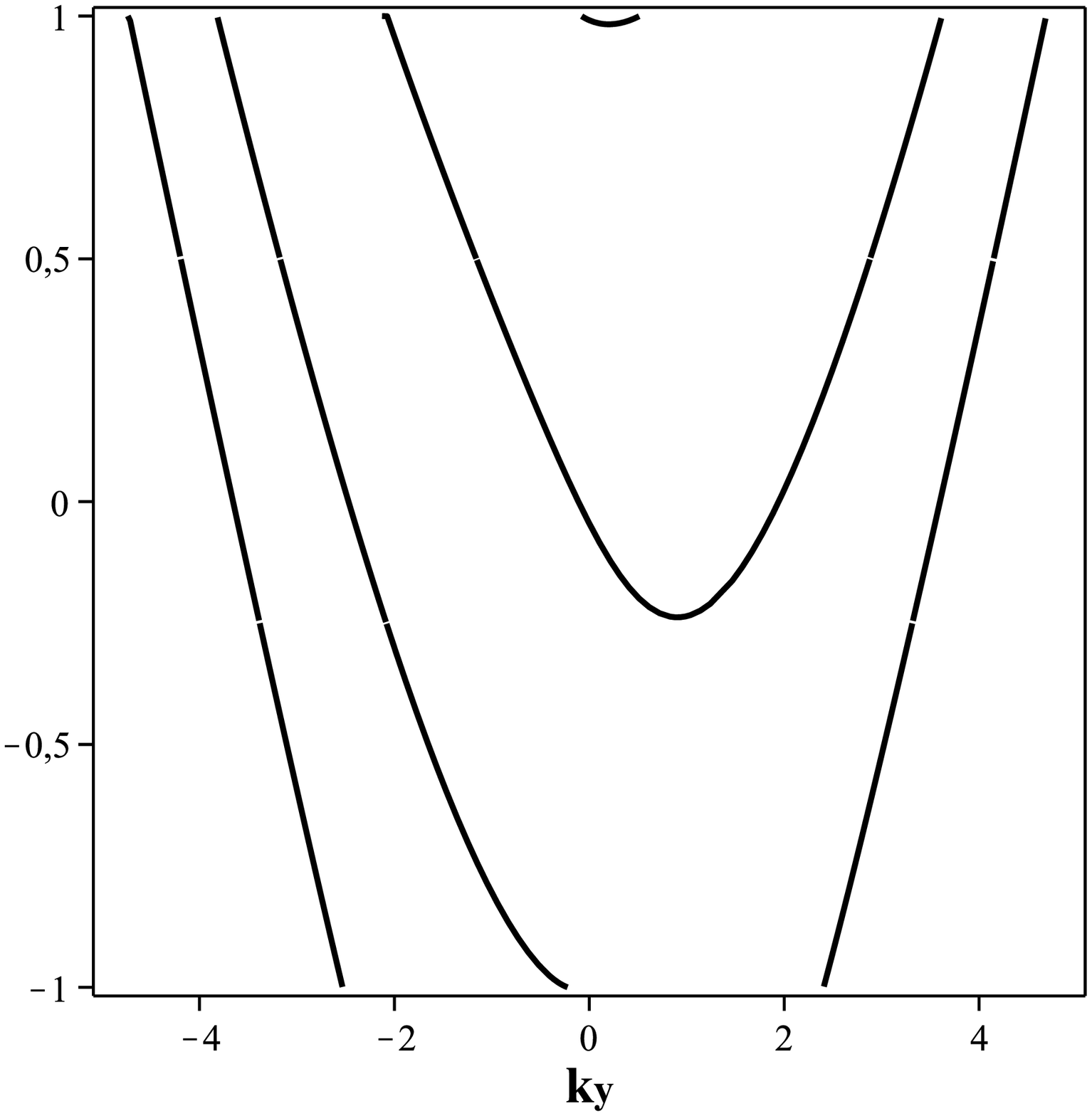}
\caption{Edge band's energies of a single TI layer  at $U_{\theta}=-4m$. Top: $L=0.4v_{\theta}/m$; bottom: $L=1.6 v_{\theta}/m$. The energy and  $k_y$ are measured in units of $m$ and $m/v_{\parallel}$, respectively}\label{fig2}.
\end{figure}

\textbf{\emph{The energy bands of a single TI layer}}. The edge states are expected to carry quantized currents in, for example, four-contact measurements of the quantum Hall effect. The number of conducting channels, however, depends on a thickness of  TI layers in a heterostructure. Fig. 2 illustrates this evident fact for a single TI layer. The single-layer states are obtained from Eq.(\ref{maineq}) by setting there $\Delta=0$. Only the states within the magnetic energy gap $-m<E<m$ are shown. Multiple bands appear at $L=1.6v_{\theta}/m$, while at $L=0.4v_{\theta}/m$ a single mode dominates in the gap. It should be noted that the number of conducting channels is always odd, so that at $m>0$ the total ideal transmission  of left moving particles exceeds that of the right movers by one. As a result, the Hall conductance is $e^2/hc$, as expected for the Chern insulator represented by Hamiltonian (\ref{H2}). In thick weakly disordered TI layers the scattering between multiple channels  will destroy this ideal picture. At the same time, in thin enough layers having only a single propagating channel the edge states stay robust. In Fig.2 (top panel) a chiral band is strongly shifted to the left from the central position $E=0, k_y=0$. Such a shift is associated with the finite potential $U_{\theta}$. In the special electron-hole symmetric case $U_{\theta}= 0$ a chiral band $E=-v_{\parallel}k_y$ can be easy obtained from Eq.(\ref{maineq}). Note that such a perfectly linear mode has been calculated on side faces of rectangular TI wires. \cite{Brey}  At $U_{\theta}= 0$ in Eq.(\ref{psis}) $k=0$ and $\phi=0$, also $\sin\phi_1=\sin\phi_2=\cos\phi_1=\cos\phi_2=1/\sqrt{2}$ in Eq.(\ref{MtMb}). The spin of the edge state on the top and bottom surfaces of the film can be calculated from Eq.(\ref{psifin}) with the factors $A$ and $B$ found from Eq.(\ref{maineq}). The spin on the side face is determined from the rotated wave-function $\mathrm{R}(\theta)\psi_{\pm}(x_{\theta})$ given by Eq.(\ref{psis}). So, we find that $\langle\sigma_x\rangle_{\text{top}}=-\langle\sigma_x\rangle_{\text{bottom}}\neq 0 $, while other spin components are zero. On the side face the spin is also precisely aligned parallel to the face. At the same time, for the modes shown in Fig.2 the spatial spin texture has not such a simple form and generally is $k_y$-dependent.

\section{Localized states}
In this section it will be shown that the Hilbert space of eigenstates of Eq. (\ref{maineq}) has a subspace of states that are localized at single TI layers and do not propagate from  layer to layer. Let us consider the range of $m>0$ and look for solutions of Eq. (\ref{maineq}) with $E=-v_{\parallel}k_y$ and $kL=\pi l$ ($l\neq 0$ is an integer). From Eq. (\ref{kphi}) it  follows that these conditions are satisfied at discrete values of $k_y$, such that
\begin{equation}\label{kl}
k_y=\frac{\pi^2l^2v_{\theta}^2}{2L^2U_{\theta}v_{\parallel}}-\frac{U_{\theta}}{2v_{\parallel}}\,.
\end{equation}
At these wave-vectors Eq. (\ref{MT}) gives $\mathrm{T}=(-1)^l$. Also, from Eqs. (\ref{kappa}-\ref{N}) and Eq. (\ref{MtMb}) we obtain at $m>\Delta$ $M_{t11}=M_{t21}=-M_{t12}=-M_{t22}=1$ and $M_{b11}=-M_{b21}=M_{b12}=-M_{b22}=1$. As a result, Eq. (\ref{maineq}) is reduced to simple equations:
\begin{equation}\label{AB2}
A_{n+1}-B_{n+1}=(-1)^l(A_{n}+B_{n})\,.
\end{equation}
A localized solution of this equation is such that for an arbitrary chosen $n^{\prime}$ we have $A_{n}=B_{n}=0$ for all $n>n^{\prime}+1$ and $n<n^{\prime}$, while $A_{n^{\prime}}=B_{n^{\prime}}=1/\sqrt{2}$ and $A_{n^{\prime}+1}=-B_{n^{\prime}+1}=(-1)^l/\sqrt{2}$. Hence, the normalized vector $\chi_n$ can be written as
\begin{align}\label{chinprime}
\chi_{n}=\frac{1}{2}
\begin{pmatrix}
1 \\
1
\end{pmatrix}
\delta_{n^{\prime},n}+\frac{(-1)^l}{2}
\begin{pmatrix}
1 \\
-1
\end{pmatrix}
\delta_{n^{\prime}+1,n}\,.
\end{align}

It should be noted that since these localized states are mostly controlled by parameters of a single TI layer, they are robust with respect to disorder, such as layer-to-layer variations in $L$, $U_{\theta}$, $\theta$, $m$ and $\Delta $. At the same time  random impurity-induced disorder can lead to scattering from a localized state to an itinerant state, if the latter exists at a given energy. Therefore, robust localized states that are isolated in the energy spectrum can exist only in structures with thin enough TI layers, as discussed in the end of Sec. II.

In a regular superlattice, the localized states are N-fold degenerate, where N is the number of TI layers in a heterostructure. It is interesting, that despite a strong tunneling, as long as $\Delta<m$, the states keep their localized character. This situation resembles the so-called flat bands in some polymer models. \cite{Su} The states, however, become itinerant at $\Delta>m$. Indeed, in this case the matrices $M_t$ and $M_b$ do not have such a special form, as in the case of $\Delta<m$. As a simple example, let us assume that $U_{\theta}=0$ and look for a solution of Eq. (\ref{maineq}) with $E=0$ and $k_y=0$. In this case from Eqs. (\ref{phi12}),  (\ref{MtMb}) and  (\ref{MT}) one obtains  Eq. (\ref{maineq}) in the form
\begin{align}\label{AB3}
\begin{pmatrix}
1&-1 \\
1& 1
\end{pmatrix}
(\chi_{n+1}-\chi_{n})=0
.
\end{align}
An itinerant solution of this equation is  $\chi_n=const$.

A delocalization of electron states at $m>\Delta$ can  be induced by a spin-dependent tunneling between TI layers.  This situation will be considered in the next section.
\section{Edge states in a regular superlattice}
In a regular superlattice $\chi_n \sim \exp(iqn)$. Therefore,  Eq. (\ref{maineq}) takes the form
\begin{equation}\label{maineq2}
\left(\mathrm{M}_s\mathrm{T}\mathrm{M}_s^{-1}\mathrm{M}_te^{iq}-
\mathrm{R}(\pi)\mathrm{M}_b\right)\chi_{n}=0\,.
\end{equation}
In the range of small energies and wave vectors $k_y$ this equation can be solved analytically.
\subsection{Analytical results}
An analytical solution of  Eq.(\ref{maineq2}) can be obtained at $E \ll |m\pm\Delta|$, $v_{\parallel}k_y \ll U_{\theta}$ and $ |m\pm\Delta|$. Depending  on a sign of $|m| - \Delta$ the system is in one of the two topologically distinct phases characterized by very different structures of eigenstates and eigenvalues of the edge bands. Accordingly, let us consider these two regimes.

\emph{\textbf{Anomalous quantum Hall regime,}  $\mathbf{|m| > \Delta}$}. In the leading approximation Eqs.(\ref{phi12}-\ref{N}) give
\begin{equation}\label{deltaphi}
\phi_{1(2)}=\text{sign}(m)\frac{\pi}{4}-\frac{\text{sign}(m)v_{\parallel} k_y +E}{2|m\pm\Delta|}\,.
\end{equation}
Further, within  this approximation and assuming  that $kL \simeq \pi l$, from Eqs.(\ref{MtMb})  and  (\ref{kphi}-\ref{MT}), one can transform  Eq.(\ref{maineq2}) to a simple form
\begin{equation}\label{maineq3}
\sin(\phi_2-\phi_1)\cos q + \cos(\phi_1 + \phi_2 +kL)=0\,.
\end{equation}
It follows from Eq.(\ref{kphi}) that for small $E$ and $k_y$ the condition $kL \simeq \pi l$ can be satisfied at $|U_{\theta}|L/v_{\theta}\simeq \pi l$. Let us write $U_{\theta}L/v_{\theta}=\pi l +\xi$, where $\xi \ll \pi$. Without loss of generality we assume that $U_{\theta}>0$. At least, this relation takes place for intrinsic potentials of non cleavage faces of Bi$_2$Se$_3$\cite{Zhang facets PRB}. Further, substituting  Eq. (\ref{deltaphi}) into  Eq. (\ref{maineq3}) and expanding the latter up to linear terms with respect to $E,\xi$ and $k_y$ we obtain the electron band energy in the form
\begin{equation}\label{Eq1}
E=\frac{\text{sign}(m)\left(\xi+\frac{v_{\parallel}}{v_{\theta}} k_yL\right)(m^2-\Delta^2)}{|m|+\frac{L}{v_{\theta}}(m^2-\Delta^2)-(-1)^l\Delta\cos q}-\frac{m}{|m|}v_{\parallel} k_y\,.
\end{equation}
It is seen that the energy is symmetric in the vertical direction, $E(q)=E(-q)$. At $\left(\xi+\frac{v_{\parallel}}{v_{\theta}} k_yL\right)=0$ the energy is independent on $q$. Hence, the flat band in the vertical direction is emerged in this parameter range. At the same time, in the $y$-direction the particle motion has a chiral character, as it is expected in the quantum Hall regime.

\emph{\textbf{Trivial phase,}  $\mathbf{|m| < \Delta}$}. A drastic reconstruction of the energy spectrum occurs upon the topological phase transition at $|m|=\Delta$. At $m<\Delta$, in the range of small $E$, $k_y$ and $kL\simeq \pi l$, instead of Eq. (\ref{deltaphi}) we obtain
\begin{equation}\label{deltaphi2}
\phi_{1(2)}=\pm\frac{\pi}{4}-\frac{\pm v_{\parallel} k_y +E}{2|m\pm\Delta|}\,.
\end{equation}
Substituting these angles into Eq. (\ref{maineq3}) we obtain at small $\left(1-(-1)^l\cos q\right)$
\begin{equation}\label{Eq2}
E=\frac{v_{\parallel} k^{\prime}_y\beta\pm\sqrt{(v_{\parallel} k^{\prime}_y)^2+2\gamma\left(1-(-1)^l\cos q\right)}}{\alpha(1-\beta^2)} +\frac{\xi\Delta}{\alpha}\,,
\end{equation}
where $\alpha=1+\Delta L/v$, $k^{\prime}_y=k_y-m\xi/\alpha v_{\parallel}$; $\beta=mL/\alpha v$ and $\gamma=(\Delta^2-m^2)(1-\beta^2)$. The two bands in Eq. (\ref{Eq2}) are represented by a combination of an anisotropic Dirac cone and a chiral part given by the first term in the numerator. The Dirac cone is placed at $E=\xi\Delta/\alpha$ and $k^{\prime}_y=0$, $q=0$ at even $l$, or $q=\pi$ at odd $l$. The chiral part results in different velocities for left and right-moving particles  (in $y$-direction). The asymmetry vanishes at $m=0$. Also, the Dirac cone becomes centered at $k_y=0$. This result is expected, because in the absence of Zeeman coupling the entire superlattice does not differ much from an anisotropic TI. Hence, the surface states which reside on the side surface are represented by conventional anisotropic Dirac bands.
\begin{figure}[tp]
\includegraphics[width=6cm]{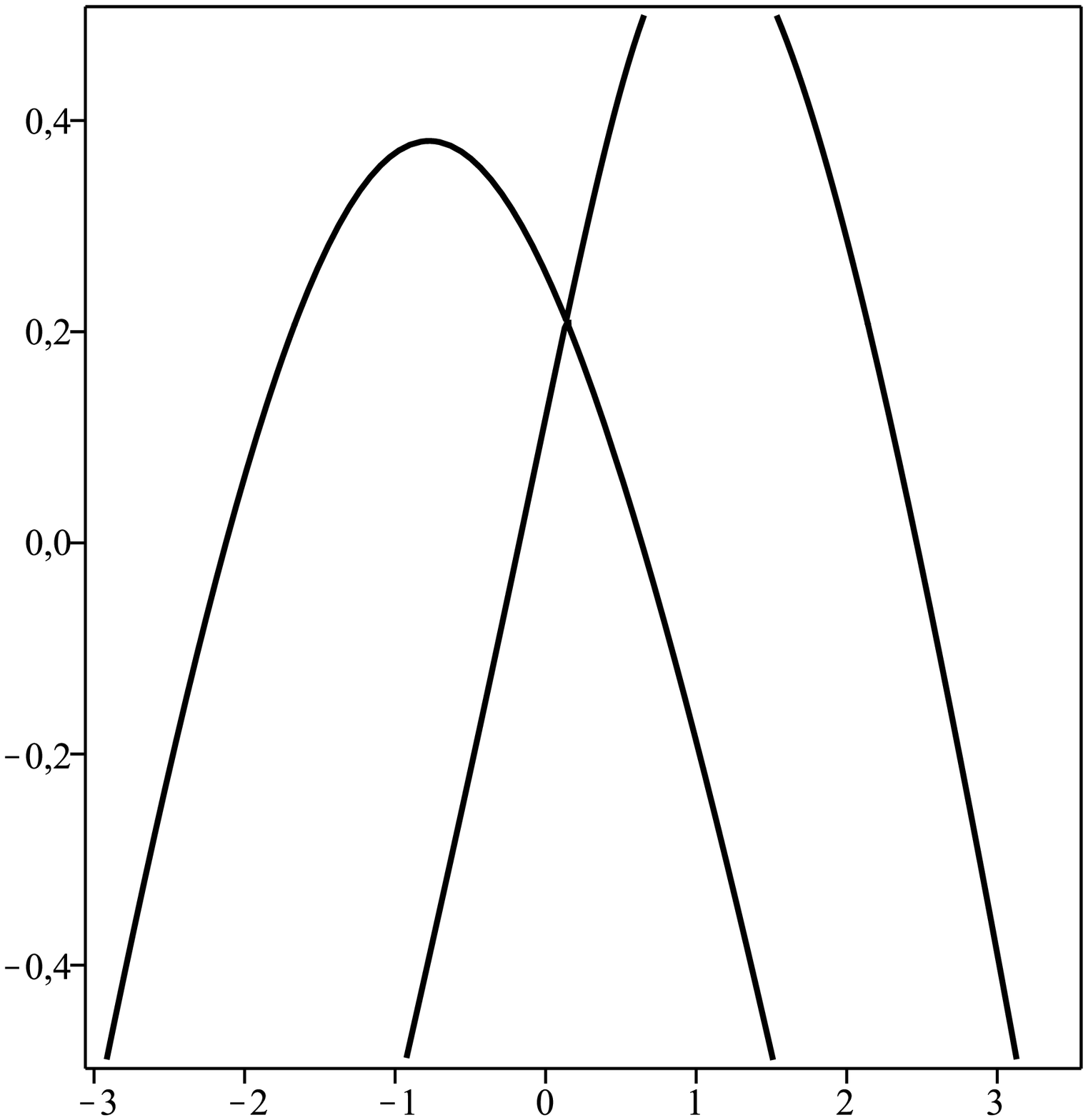}
\includegraphics[width=6cm]{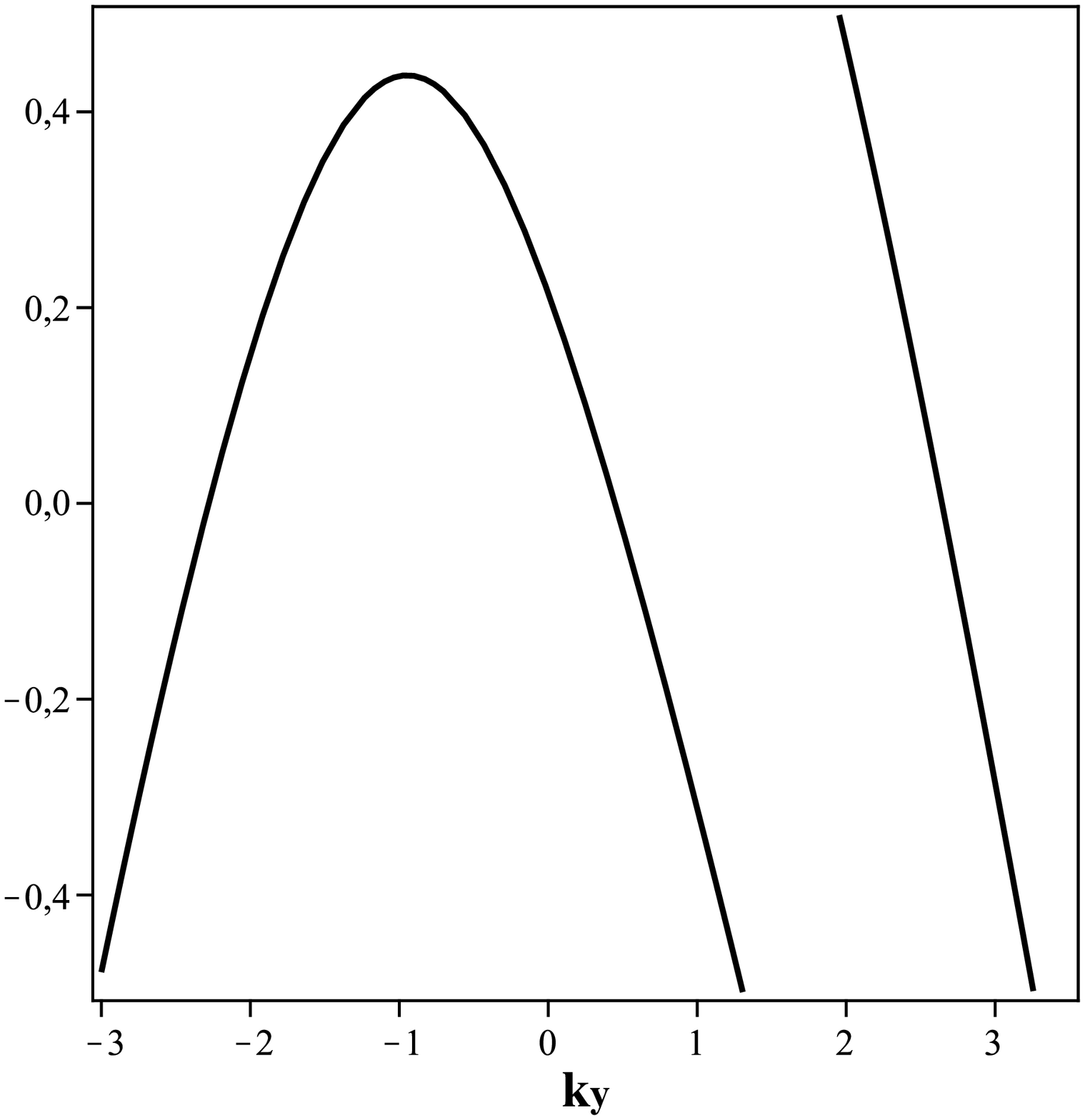}
\caption{Edge band's energies of a superlattice at $q=-\pi$, $U_{\theta}=3.5m$ and $L=1v_{\theta}/m$. Top: $\Delta=1.5m$; bottom: $\Delta=0.5m$. }\label{fig3}.
\end{figure}

\emph{\textbf{Spin-dependent tunneling}}. Now let us take into account a possible dependence of the tunneling amplitude on the particle spin.  As an example, let us assume that in addition to $\Delta\tau_1$, Hamiltonian  (\ref{H}) contains the spin-dependent tunneling  $\Delta_s\tau_1\sigma_z$. Such a spin dependence may be associated with a magnetic ordering in the tunnel barrier.   In unitary transformed Hamiltonian  (\ref{H2}) this term takes the form $\Delta_s\tau_3$. As a result,  Eqs. (\ref{kappa}-\ref{phi12}) are modified, so that $E$ is substituted for $E\pm \Delta_s$. An immediate effect of this modification is that at $|m|>\Delta$  Eq.(\ref{Eq1}) takes the form
\begin{equation}\label{Eq3}
E=E_0+\frac{\gamma((-1)^lm\cos q-\frac{m}{|m|}\Delta)}{|m|+\frac{L}{v_{\theta}}(m^2-\Delta^2)-(-1)^l\Delta\cos q}\,,
\end{equation}
where $E_0$ is given by Eq.(\ref{Eq1}). It is seen that at finite $\gamma$ the edge-state energy depends on $q$ in the whole parameter range and the flat band regime can not be reached.
\subsection{Numerical results}
\begin{figure}[tp]
\includegraphics[width=6cm]{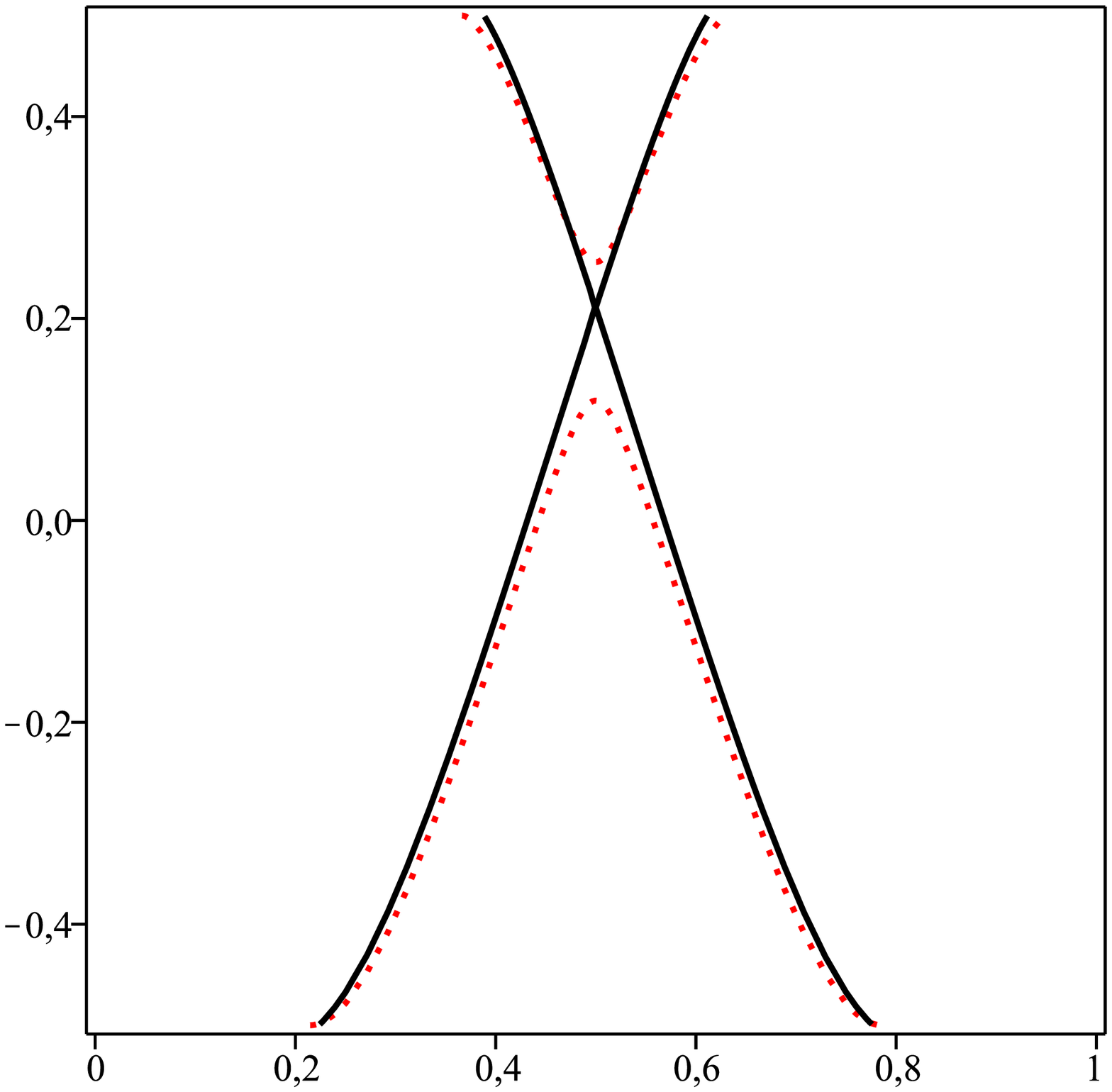}
\includegraphics[width=6cm]{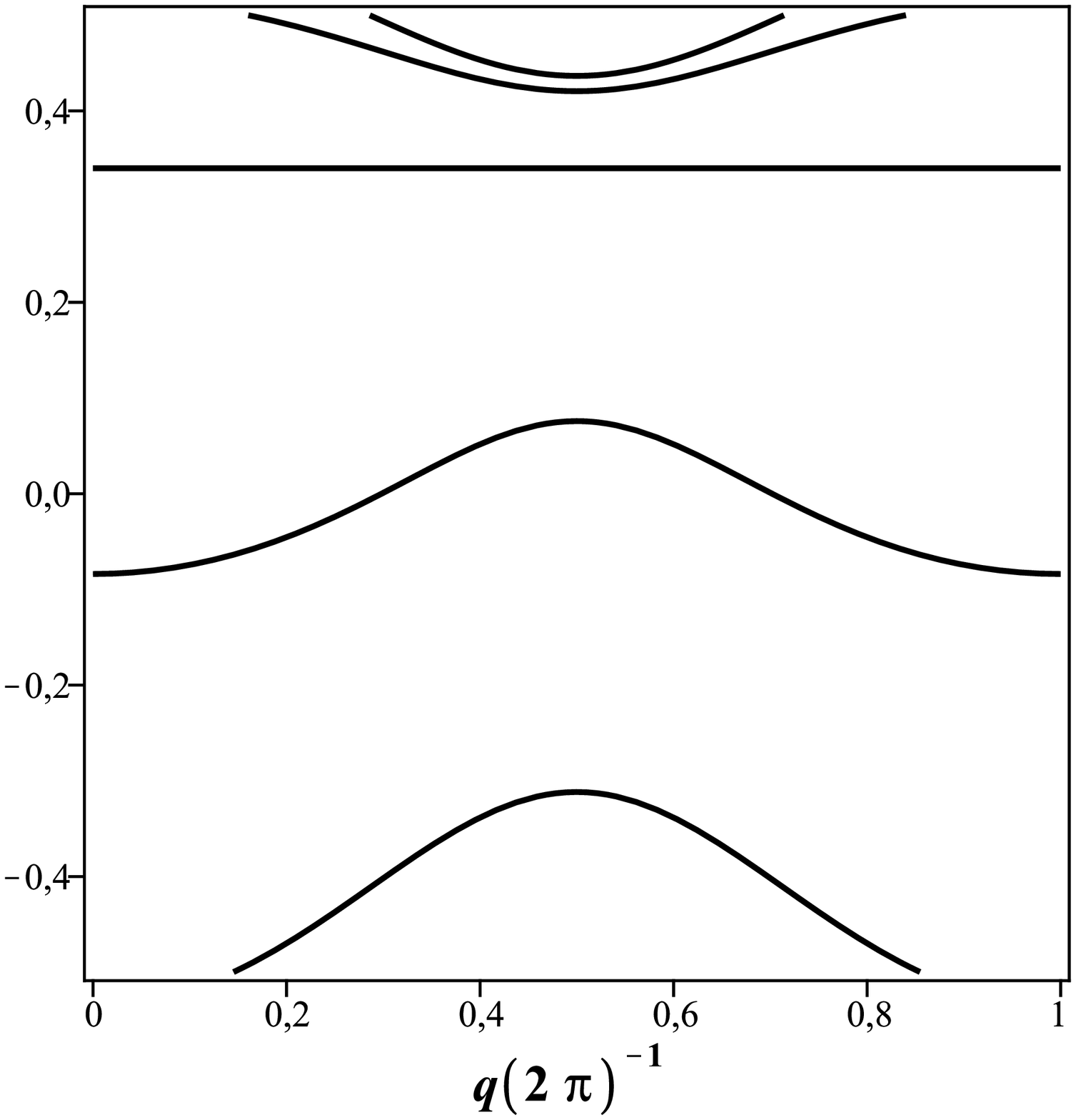}
\caption{(Color online) Edge band's energies of a superlattice, as a function of $q$ at  $U_{\theta}=3.5m$ ,$L=1v_{\theta}/m$. Top: $\Delta=1.5m$, $k_y=0$ (dotted), $k_y=0.138$ (solid); bottom: $\Delta=0.5m$, $k_y$ takes the values (from top curve to bottom curve) $k_y=-1.3,-0.7,-0.34,0.3,1$.}\label{fig4}.
\end{figure}

Two examples of edge-bands at fixed $q=\pi$ and varying $k_y$ are shown in Fig.3. The band energies are shown within the gap $|m-\Delta|$ for  $m>0$, the energy and  $k_y$ are measured in units of $m$  and $m/v_{\parallel}$, respectively. The bottom plot in Fig.3 corresponds to the quantum Hall regime $\Delta < m$. Similar to Fig.2, one can see the odd number of propagating channels at each energy, so that the number of left moving particles is larger by one channel. In contrast, in a trivial phase at $\Delta > m$ (top plot) the number of channels is even and a Dirac point appears at, approximately, $k_y=0$ and $E=0.2m$.

The bands at the varying wavenumber $q$ in vertical direction, at fixed $k_y$ are shown in Fig.4. A Dirac point at $k_y=0.138$ is seen in the top plot (solid line) calculated at $\Delta>m$. The Dirac point vanishes in the quantum Hall regime shown in the bottom plot. At the same time, in this regime the flat band appears at $k_y=-0.34 m/v_{\parallel}$ and $E=0.34m$, as it follows from Eq. (\ref{kl}) for $l=1$. This band of states localized in the $z$-direction resides on the parabolic-like dispersion curve in the bottom plot of Fig. 3. As one can see from this plot, there are two more modes having the same energy as the localized state has. Therefore, this mode is not protected from an impurity scattering causing transitions between the localized and itinerant modes. An inspection of energy bands in a broad range of parameters has shown that, probably, there are no isolated flat bands and that they always coexist with other modes at the same energy. In this moment it is difficult to say whether this fact is general, or it is associated with the specific model studied here.
\section{Discussion}
In this work the energy bands of electrons propagating on the side surface of a heterostructure which consists of TI and magnetic insulator layers has been studied. We found out that stable with respect to weak disorder isolated chiral bands, which could serve as carriers of a quantized Hall signal, can exist only on the surface of a heterostructure having thin enough ($L\lesssim v_{\theta}/m$) TI films.  Such bands have been calculated in the anomalous quantum-Hall regime $|m|>\Delta$. In contrast, on the other side of the topological phase transition, at $|m|<\Delta$ the energy bands have a quite different structure that resembles more the anisotropic Dirac bands which exist on the side surface of an anisotropic TI crystal.

We found out that a regular, as well as irregular in the z-direction heterostructure, carries the states that are localized at the vicinity of each TI layer and are characterized by a specific for each state wave-vector in the horizontal direction. In the regular case these states form a flat band in the vertical direction. An interesting property of such localized states is that they stay localized even in the case of a strong tunneling between TI layers. Their localization has a topological nature and is associated with their specific spin texture. Namely, electron spins on the top and bottom surfaces of a TI layers are oppositely oriented. Therefore the tunneling is blocked, as far as spin-flip tunneling processes are weak, as shown  in Sec. IV.  The spin-dependent tunneling can be considerably suppressed if, instead of magnetic layers, the nonmagnetic layers are incorporated  into the heterostructure. In this case the magnetic order and Zeeman coupling in Eq. (\ref{H}) could be provided by magnetic impurities in TI bulk. \cite{magnetic  doping}  Besides the spin-dependent tunneling, the localized states can be destroyed by a strong impurity scattering causing electron transitions between a localized mode that exists at some value of the horizontal wave-vector $k_y$ and itinerant modes (if there are any) at other wave-vectors . If the disorder is weak, the localized levels become broadened, so that a considerable density of such states can present in a narrow energy interval. This narrow band may show up in transport measurements. For example, in the four-contact measurements of the Hall effect a pair of contacts could be attached to one of the several TI layers, while the other pair could be connected to the same layer, or to another layer in a heterostructure. It is interesting to measure difference in Hall responses of these two setups when the Fermi level is tuned into the band of localized states. In this case a sharp increase of the Hall signal should be expected in the first set-up, even if  the tunneling parameter $\Delta$ is comparable to $m$.

Besides transport, flat bands can influence strongly on various phenomena involving electron-electron, or electron-phonon interactions, such as surface plasmons, Wigner crystallization and superconductivity.

The localized states, as well as the Hall response are expected to vanish at $\Delta>|m|$. At the same time, the topological phase transition at $\Delta=|m|$ can be realized by varying, for example, the temperature near the Curie point.



\begin{thebibliography}{99}
\bibitem{Qi RMP}
X. L. Qi and S. C. Zhang, Rev. Mod. Phys. {\bf 83}, 1057
(2011).
\bibitem{Hasan}
M. Z. Hasan, C. L. Kane, Rev. Mod. Phys. {\bf 82}, 3045 (2010)
\bibitem{Fu}
L. C. Fu, C. L. Kane, E. J. Mele, Phys. Rev. Lett. {\bf 398}, 106803 (2007)
\bibitem{magnetic doping}
Y. L. Chen, {\it et. al.}, Science {\bf 329}, 659 (2010); S.-Y. Xu, {\it et. al.}, e-print arXiv:1206.2090; M. Liu, {\it et. al.}, Phys. Rev. Lett. {\bf 108}, 036805 (2012); Y. S. Hor, {\it et. al.}, Phys. Rev. B {\bf 81}, 195203 (2010); C. Z. Chang, {\it et. al.}, Science {\bf 340}, 167 (2013).
\bibitem{Wei}
P. Wei, F. Katmis, B. A. Assaf, H. Steinberg, P. Jarillo-Herrero, D. Heiman, and J. S. Moodera, Phys. Rev. Lett. {\bf 110}, 186807 (2013); A. Kandala, A. Richardella, D. W. Rench, D. M. Zhang, T. C. Flanagan, and N. Samarth, Appl. Phys. Lett. {\bf 103}, 202409 (2013)
\bibitem{Qi Basics TI}
 X. L. Qi, T. Hughes, and S. C. Zhang, Phys. Rev. B {\bf 78}, 195424 (2008)
\bibitem{Essin}
 A. M. Essin, J. E. Moore, and D. Vanderbilt, Phys. Rev. Lett. {\bf 102}, 146805 (2009)
\bibitem{TME}
X. L. Qi, R. Li, J. Zang, and S. C. Zhang, Science {\bf 323}, 1184 (2009); K. Nomura and N. Nagaosa, Phys. Rev. B {\bf 82}, 161401(R)  (2010); G. Rosenberg, H.-M. Guo, and M. Franz, Phys. Rev. B {\bf 82}, 041104 (2010).
\bibitem{Zhang facets}
Fan Zhang, C. L. Kane, and E. J. Mele, Phys. Rev. Lett. {\bf 110}, 046404 (2013)

\bibitem{Meng folding}
Q. Meng, S. Vishveshwara, and T. L. Hughes, Phys. Rev. Lett. {\bf 109}, 176803 (2012)
\bibitem{calculations for superlattices}
Weidong Luo and Xiao-Liang Qi, arXiv: 1208.4638
\bibitem{suplattice Bi2Bi2Se3}
T. Valla, Huiwen Ji, L. M. Schoop, A. P. Weber, Z.-H. Pan, J. T. Sadowski, E. Vescovo, A. V. Fedorov, A. N. Caruso, Q. D. Gibson, L. M\"{u}chler, C. Felser and R. J. Cava, Phys. Rev. B {\bf 86}, 241101(R) (2012)
\bibitem{natural heterostruc Ando}
K. Nakayama, K. Eto, Y. Tanaka, T. Sato, S. Souma, T. Takahashi, Kouji Segawa, and Yoichi Ando, Phys. Rev. Lett. {\bf 109}, 236804 (2012)
\bibitem{Ji intergrowth}
H. Ji, J. M. Allred, N. Ni, J. Tao, M. Neupane, A. Wray, S. Xu, M. Z. Hasan, and R. J. Cava, Phys. Rev. B {\bf 85}, 165313 (2012)


\bibitem{Brey}
L. Brey and H.A. Fertig,  e-print arXiv:1312.5593
\bibitem{Deb}
O. Deb, A. Soori, D. Sen, e-print arXiv:1401.1027
\bibitem{Su}
W. P. Su, J. R. Schrieffer, and A. J. Heeger, Phys. Rev. Lett. {\bf 42}, 1698 (1979).
\bibitem{Zhang facets PRB}
Fan Zhang, C. L. Kane, and E. J. Mele, Phys. Rev. B {\bf 86}, 081303(R) (2013)
\bibitem{Silvestrov}
P. G. Silvestrov, P. W. Brower, and E. G. Mischenko, Phys. Rev. B {\bf 86}, 075302 (2012)
\bibitem{redefine}
Compare to Ref. \onlinecite{Zhang facets PRB} the pseudospin $\bm{\tau}$ is $\pi/2$ rotated around the $x_2$-axis, so that $\tau_1 \rightarrow -\tau_3$ and $\tau_3 \rightarrow \tau_1$
\bibitem{Eremeev}
S. V. Eremeev, V. N. Men'shov, V. V. Tugushev, P. M. Echenique, and E. V. Chulkov, Phys. Rev. B {\bf 88}, 144430 (2013); S. V. Eremeev, V. N. Men'shov, V. V. Tugushev, P. M. Echenique, and E. V. Chulcov, Phys. Rev. B {\bf 88}, 144430 (2013); V. N. Men'shov, V. V. Tugushev, S. V. Eremeev, P. M. Echenique, and E. V. Chulkov, Phys. Rev. B {\bf 88}, 224401 (2013)
\bibitem{film Hamiltonian}
H. Z. Lu , W. Y. Shan , W. Yao, Q. Niu  and S. Q. Shen , Phys. Rev. B {\bf 81}, 115407 ( 2010); J. Linder, T. Yokoyama, and A. Sudb{\o}, Phys. Rev. B,
{\bf 81}, 205401 (2009); C.-X. Liu, H. J. Zhang, B. Yan, X.-L. Qi, T. Frauenheim, X. Dai, Z. Fang, and S.-C. Zhang, Phys. Rev. B {\bf 81}, 041307 (2010).



\end{thebibliography}
\end{document}